\documentclass[aps,prb,twocolumn,showpacs,preprintnumbers,amsmath,amssymb]{revtex4-1}

\usepackage{graphicx}
\usepackage{dcolumn}
\usepackage{bm}
\usepackage{natbib}
\usepackage{amsmath}
\usepackage{braket}
\usepackage{mathrsfs}
\usepackage{amsfonts}
\usepackage[margin=1in]{geometry} 
\usepackage{times}
\usepackage{xcolor}

\begin{document}

\title{Spin decoherence in {\it n}-type GaAs: the effectiveness of  the third-body rejection method for electron-electron scattering}

\author{Gionni Marchetti} 
\email{gionnimarchetti@gmail.com}
\author{Matthew Hodgson}
\email{matthew.hodgson@york.ac.uk}
\author{Irene D'Amico}
\email{irene.damico@york.ac.uk}
 
\affiliation{
Department of Physics, University of York, York, Heslington YO10 5DD, UK\\
}

\date{\today}
\begin{abstract}

We study the spin decoherence in n-type bulk GaAs for moderate electronic densities at room
temperature using the Ensemble Monte Carlo method. We demonstrate that a technique called
 \textquotedblleft third-body rejection method\textquotedblright, devised by Ridley [B.~K.~Ridley, J. Phys. C: Solid State Phys. \textbf{10}, 1589, 1977]
 can be successfully adapted to ensemble Monte Carlo method and used to tackle the problem of the electron-electron contribution to spin decoherence 
 in the parameter region under study, where the electron-electron interaction can be reasonably described by a
Yukawa potential. This  scattering technique is employed in a doping region where one
can expect that multiple collisions may play a role in carrier dynamics. By this technique we are
able to calculate  spin relaxation times which are in very good agreement with the experimental
results found by Oertel et al.[S.~Oertel, J.~H\"{u}bner, M. Oestreich, Appl. Phys. Lett., \textbf{93}, 13, 2008]. Through this method we show
 that the electron-electron scattering is overstimated in Born approximation, in agreement with previous results
obtained by Kukkonen and Smith [C.~A.~Kukkonen, H.~Smith, Phys. Rev. B, \textbf{8}, 4601, 1973].
\end{abstract}

\maketitle

\section{Introduction}\label{sec:Introduction}

Spintronics is a highly active field that encompasses both fundamental research  and practical applications. Its goal is to study 
and exploit spin-related properties  in material, e.g. in metals, semiconductors and semiconductor heterostructures, 
as well as in more exotic structures such as topological insulators or organic molecules\cite{bader2010,kusrayev2010}. 
The coherent transport of spin is a central issue in spintronics. For this reason, the spin transport  in
{\it n}-type GaAs with its long electronic spin  lifetime \cite{kikkawa1998}, has been recently studied both theoretically 
\cite{glazov2002,glazov2003,jiang2009} and experimentally \cite{oertel2008,romer2010}. An important goal is to 
achieve a clear understanding of spin decoherence phenomena due to the  carrier dynamics, and  in different doping regimes. 

In bulk n-GaAs at high temperatures and for the range of doping densities here considered ($n_{\mathrm{e}}$ from $10^{16}$ $ \mathrm{cm}^{-3}$ to $2.5\cdot 10^{17}$ $ \mathrm{cm}^{-3}$), 
the main source of spin relaxation is the Dyakonov-Perel (DP) mechanism \cite{dyakonov2008},  a type of spin-orbit interaction. This mechanism arises 
from the bulk inversion asymmetry \cite{dresselhaus2008} giving rise to effective, momentum-dependent magnetic fields. Thus due to 
collisional events which cause momentum transfer, each electronic spin
undergoes a precession around a different direction. Such a kinematics gives
rise to spin dephasing. 

In previous works \cite{sheetal2011, marchetti2014} we studied  spin transport using the Ensemble Monte Carlo (EMC) method. 
EMC is a stochastic method devised to solve numerically the Boltzmann equation for charge transport in semiconductors \cite{jacoboni1989, fischetti1998} which
is also suitable for studying  spin dynamics  \cite{sheetal2011,saikin2005}. It is worthwhile to recall here that the EMC method applied
 to charge transport has provided very accurate estimates of semiconductor material properties, e.g. drift velocities, 
electron mobilities, etc \cite{jacoboni1989}. However the electron-electron (e-e) interactions do not affect greatly the charge transport calculations, 
due to the conservation of the total energy and momentum, so they could be generally discarded in the EMC calculations \cite{jacoboni1989, combescot1988}.
This is not the case for spin dephasing where the e-e interactions play an important role as theoretically predicted, e.g. by Refs. \onlinecite{glazov2002,glazov2003,jiang2009}.

In Ref.~\onlinecite{marchetti2014} we simulated the effect of electron-electron  scattering on the spin relaxation time (SRT). We found that
the inclusion of the e-e scattering was the key to reproduce the experimental results found by Oertel et al. \cite{oertel2008}.
Our results for {\it n}-type bulk GaAs at relatively high temperatures ($280~ \mathrm{K} \le \mathrm{T}\le 400~\mathrm{K}$) and moderate doping concentrations 
($n_{\mathrm{e}}$ from $10^{16}$ to $10^{17}$ $ \mathrm{cm}^{-3}$) were in very good agreement with  experiments.

However,  for electronic densities  $n_{\mathrm{e}} \stackrel{>}{\sim} 10^{17}  \mathrm{cm}^{-3}$ our calculations overstimated the spin relaxation time.
We interpreted this  as an effect of the failing of the first order Born approximation (BA) (usually simply referred to as \textquotedblleft Born
Approximation \textquotedblright) for the electron-electron scattering. In fact, the Ensemble Monte Carlo algorithm relies on the Fermi Golden Rule 
which entails the Born approximation, and the latter is well known to  perform less well as an approximation for low energy collisions \cite{kukkonen1973,joachain1987}. 
Our analysis of the e-e scattering contribution to the spin relaxation time \cite{marchetti2014}, indicates that the electron-
electron scattering is overestimated. This result depends also on the interelectronic potential  adopted in the calculations, modeling
the conduction electrons as an electron gas (jellium model). In the parameter region  where the random phase approximation (RPA) holds,
a Yukawa potential can be used \cite{giuliani2005}. It is reasonable to assume that for $n_{\mathrm{e}} \stackrel{>}{\sim} 10^{17}  \mathrm{cm}^{-3}$  and at 
 temperature  $\mathrm{T}=300$ $\mathrm{K}$ a  Yukawa potential is adequate to picture the screened short-range Coulomb interaction. 

In Ref.~\onlinecite{marchetti2014} within this model we observed that the discrepancies between the calculations  and the experimental results in Ref.~\onlinecite{oertel2008}
become larger when the electronic density increases.  In this circumstance the average distance between 
electrons in the conduction band decreases and thus there are reasons for believing that scattering processes which involve more 
than two electrons simultaneously may become important. 
Since the scattering formulae usually used in EMC simulations, see Ref.~\onlinecite{jacoboni1989}, rely on the assumption of truly two-body processes, 
in order to improve the agrement with the experimental results it is necessary to include this possible
physical effect in the calculations. Here we propose to use a simple method called \textquotedblleft third-body rejection \textquotedblright \, (tbr) 
introduced by Ridley \cite{ridley1977,ridley2000}.
This method takes into account the potential third body in a multiple scattering event by introducing an additional probability factor - the
\textquotedblleft third-body exclusion factor \textquotedblright \, - which, by renormalizing the 
scattering cross-section, assures that the collisions are truly two-body processes.

We find that this method  improves our SRT calculations at higher  doping densities, bringing  very good agreement 
with the experimental results and therefore significantly improving over  Ref.~\onlinecite{marchetti2014}.

\section{Ensemble Monte Carlo Method}\label{sec:Methods}

In this section we shall give a brief account of the computational aspects of our study. For more detail we refer to Ref.~\onlinecite{marchetti2014}.

The Ensemble Monte Carlo method is a semiclassical numerical approach suitable to describe  charge and spin
transport  in semiconductors. It includes a sequence of free flights for each simulated particle  whose durations are randomly
generated. Each free flight time is terminated by  a scattering
event \cite{jacoboni1989}. After the collision, the energy and momentum of the particle are updated according to  one of
the possible scattering mechanisms. The process is repeated until enough data are generated according to the aims
of the simulation.

Each free flight time $\tau_{0}$ is given by \cite{jacoboni1989}
\begin{equation}
\tau_{0} = −\ln(r)/\Gamma_{\mathrm{tot}}
\end{equation}
where $r$ is a random number generated stochastically  from a uniform distribution on the interval $(0,1)$ and $\Gamma_{\mathrm{tot}}$ is the total
scattering rate which includes all the scattering mechanisms of the system under  study. $\Gamma_{\mathrm{tot}}$ is calculated at 
the beginning of the simulation as a function of the colliding particle energy \cite{marchetti2014}. In between the scattering events, 
 carriers propagate along a classical trajectory which may be influenced by external forces
due to applied electric and/or magnetic fields. In the present work we simulate the dynamics of $N=25,000$ carriers.

EMC allows us to follow the spin dynamics together with the carrier dynamics. 
During the carriers' free flights the electron spins are considered non-interacting, and each spin undergoes a coherent evolution dictated by
the spin-dependent part of the  Hamiltonian \cite{marchetti2014}.  For the system under consideration, the main source 
of spin relaxation is spin-orbit interaction due to the bulk inversion asymmetry. This is described by the Dresselhaus Hamiltonian $H_{\mathrm{D}}$ 
\cite{fabian2007,dresselhaus2008}
\begin{equation}\label{eq:dresselhaus}
H_{\mathrm{D}}=\hbar  \mathbf{\Omega}(\mathbf{k})\cdot \vec{\mathbf{\sigma}} \, ,
\end{equation}
where $\vec{\mathbf{\sigma}}=\left(\sigma_x,\sigma_y,\sigma_z\right)$ are the Pauli matrices, and the Larmor precession frequency vector $\mathbf{\Omega}(\mathbf{k})$ is
\begin{equation}
 \mathbf{\Omega}(\mathbf{k})=\frac{\gamma_{so}}{\hbar}[k_{x}(k_{y}^2-k_{z}^2), k_{y}(k_{z}^2-k_{x}^2), k_{z}(k_{x}^2-k_{y}^2)] \, .
\end{equation}
Here $k_{i}$  are the wavevector components along the cubic crystal axes, $i=x,y,z$, and  $\gamma_{so}$ is known as the Dresselhaus coefficient,
whose values are determined using different methods. In GaAs, $\gamma_{so}$ values 
have been suggested which range from $8.5$ to $34.5$ $\mathrm{eV}$ \AA $^{3}$ \,   \cite{fu2008}.

We are interested in studying the system at equilibrium, therefore  we let the system thermalise for a suitable time \cite{marchetti2014}, after which 
we realign the electron spins along  the $z$-axis. Afterwards the spins will  dephase  via
the Dyakonov-Perel mechanism whereby each spinor wavefunction is acted upon by the time evolution operator generated by
the  Hamiltonian $H_{\mathrm{D}}$.

Using  EMC we can study the spin dynamics
of each carrier from its spinor wavefunction. At any given time we can extract the
expectation values of the $S_x$, $S_y$ and $S_z$ components of the individual electron spin operator $S$ to get the
probability for the spin to be aligned along each direction. Because we  start from an electronic ensemble fully polarized along the $z$ -axis,
we focus on the time evolution of the expectation value of the total  $z$-component spin operator $S_{z,\mathrm{tot}}$. For each simulation, by plotting 
$S_{z,\mathrm{tot}}$ against time, and assuming an exponential behaviour, we fit  the  data from the simulation 
and extract the corresponded spin relaxation time $\tau_\mathrm{s}$ \cite{marchetti2014}.

\subsection{Scattering Types And Related Approximations}\label{sec:model}

We consider spin transport in bulk GaAs at $\mathrm{T}=300$ $\mathrm{K}$. The carriers' dynamics are described in the framework of a 
single parabolic energy  band (the central $\mathrm{\Gamma}$ valley) which gives rise to an
effective isotropic electron mass $m^{\ast}=0.067 m_{\mathrm{e}}$ where $m_{\mathrm{e}}$ is the bare electron mass. We do not include the 
valleys $\mathrm{X}$ and $\mathrm{L}$  because the electrons we simulate have  energies which give a negligible probability to scatter into these valleys \cite{yu2010}.

The carriers (electrons) in our simulations undergo scattering with longitudinal acoustic (LA)  phonons,  polar 
longitudinal optical (LO) phonons, singly-ionized impurities in Brooks-Herring approach \cite{jacoboni1989}  
and finally with other electrons. Electron-piezoacoustical interactions are not included because 
they become relevant for GaAs samples only at low temperatures \cite{ridley2000}. The scattering rate for the electron-LA phonon collisions
is determined by the acoustic deformation potential in elastic approximation, as inelastic absorption/emission processes are  important only 
at low temperatures \cite{jacoboni1989}. Due to the space group selection rules, for the electrons in the $\mathrm{\Gamma}$ valley there is 
no deformation potential interaction with the optical phonons \cite{yu2010}.
The electron-LO scattering rate (Fr\"{o}hlich interaction \cite{mahan2000}) includes absorption  and 
emission processes with a threshold energy of $35$ meV. Phonons are considered at equilibrium at the lattice temperature $\mathrm{T}$. 
We use the  Fermi Golden rule to calculate  the scattering rates at first order for each type of  scattering mechanism.

We work within the random phase approximation \cite{giuliani2005}, which neglects the exchange and correlation effects: we model  
the screened Coulomb interaction between two  charges as in a homogeneous electron gas using the (Yukawa-type) Coulomb 
potential
\begin{equation}\label{eq:ee-potential}
  V\left(r\right)=\frac{\mathrm{e}^{2}}{4 \pi \varepsilon r}
  e^{-\beta_{\mathrm{TF}} r } \, .
\end{equation}
Here $r=|\mathbf{r}_{A}-\mathbf{r}_{B}|$ is the distance between the interacting electrons denoted by $A$ and $B$, and 
$\varepsilon$ is the material dielectric constant, $\varepsilon=12.9  \varepsilon_{0}$ in the case of GaAs. The quantity
$\beta_{\mathrm{TF}}$ is called Thomas-Fermi wavevector or \textquotedblleft inverse screening length\textquotedblright \, and is derived in the framework
of a finite temperature linearized Thomas-Fermi approximation (LTFA)\cite{marchetti2014,sanborn1995}. For {\it n}-type semiconductors
with a parabolic band, it is given by the following expression
\cite{chattopadhyay1981, dingle1955} 
 
\begin{equation}\label{eq:screening}
\beta_{\mathrm{TF}}^{2}=\frac{n_{\mathrm{e}} \mathrm{e}^{2}}{\varepsilon k_{\mathrm{B}}\mathrm{T}}\frac{\mathscr{F}_{-1/2}(\eta)}{\mathscr{F}_{1/2}(\eta)} \, . 
\end{equation}

Here $\mathrm{e}$ and $k_{\mathrm{B}}$  are the electron charge and the Boltzmann constant respectively,  $\mathscr{F}_{j}$ denotes the Fermi-Dirac integral 
of order $j$, $\eta=\mu/\left(k_{\mathrm{B}}\mathrm{T}\right)$ and $\mu$ is the electronic chemical potential. If the RPA holds, then the LTFA is a fair 
approximation only insofar  as we have small disturbances in the electron gas, or equivalently small momentum transfer in Coulomb
scattering processes, see Refs.~\onlinecite{ashcroft1976,marchetti2014}.
 
For completeness we should also mention  that in the (homogeneous) electron gas model proposed by Pines and Bohm's theory 
\cite{pines1952,lugli1983,lugli1985}the Coulomb interaction is split in two contributions:  a Yukawa Coulomb scattering between individual 
electrons and an electron-plasmon interaction.
The latter  is not included in our calculations because the electron-plasmon scattering becomes important in GaAs 
for higher electronic concentrations than considered in the present work \cite{jacoboni1989}.

In order to make the e-e scattering  consistent with the Pines and Bohm's model described above,
when a carrier undergoes an e-e collision in our EMC simulations, its electron partner is chosen arbitrarly but 
within a distance of one screening length. This is an improvement over commonly used algorithms which select the second electron 
from the whole ensemble and with a uniform distribution. Our choice describes much better the  locality of a 
screened electron-electron interaction. Moreover this is consistent with the e-e scattering rate we 
employ as it follows by assuming a local field theory \cite{peskin1995}.

\begin{figure}

\begin{center}
\includegraphics*[scale=0.6]{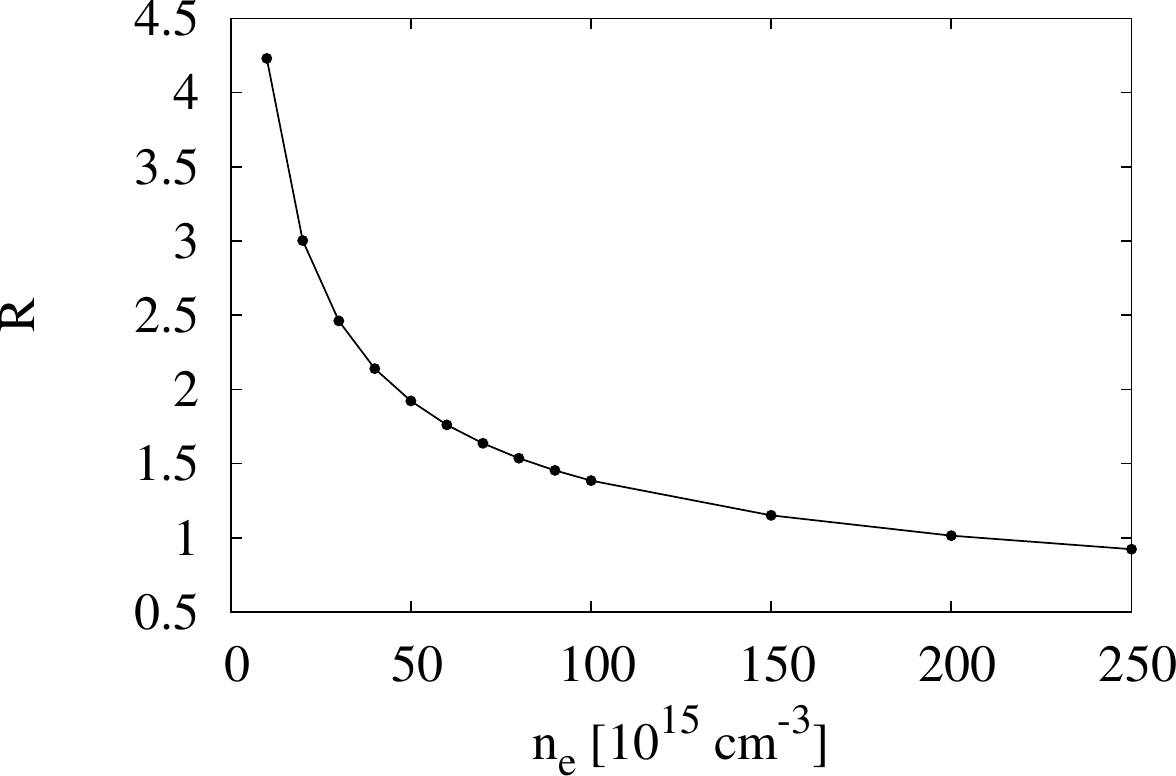}
\end{center}

\caption[] {The inequality given by Eq. (\ref{eq:born}) against the electron density at $\mathrm{T}=300$ $\mathrm{K}$. }
\label{fig:born}
\end{figure}

Finally we shall say a few words about the e-e scattering for low energy collisions.
In the low energy limit the first order Born approximation for a potential given by  Eq. (\ref{eq:ee-potential})  is valid when \cite{schiff1968}
\begin{equation}\label{eq:born}
 \mathrm{R}=\frac{m^{\ast}  \mathrm{e}^{2} \lambda_{\mathrm{TF}}}{4 \pi \varepsilon \hbar^{2}}=\frac{ \lambda_{\mathrm{TF}}}{a^\ast_{\mathrm{B}}} \ll 1 \, ,
\end{equation}
with $a^\ast_{\mathrm{B}}=(4\pi\hbar^{2} \varepsilon )/(\mathrm{e}^{2}m^{\ast})$, the effective Bohr radius and $\lambda_{\mathrm{TF}}=1/\beta_{\mathrm{TF}}$. The inequality (\ref{eq:born}) is 
not satisfied  for the range of densities considered here, as $\mathrm{R}\stackrel{<}{\sim} 1$ (see Fig. \ref{fig:born}). This confirms the 
fact that in general the BA  is not adequate for low energy processes \cite{joachain1987,ziman1972}.

By using the  phase-shift calculation of electron-electron scattering \cite{moskova2000}, Kukkonen and Smith\cite{kukkonen1973}   have
found that the electron-electron total cross-section  in BA  for  a metal like 
Na is overstimated by a factor two. Kukkonen and Smith
 assumed a scattering potential as in Eq. (\ref{eq:ee-potential}) and included also the antisymmetry of the wavefunction of 
the colliding carriers.  In the case of an electronic gas in a solid, the average interelectronic distance $r_{\mathrm{s}}$ is defined 
by the relation \cite{ashcroft1976}
$n_{\mathrm{e}}^{-1}=(4 \pi/3)\left(r_s a^\ast_{\mathrm{B}} \right)^{3}$; for Na its value
is $3.96$. Therefore we might expect a similar trend in  semiconductors with a comparable $r_{\mathrm{s}}$, as in the case here considered.

\section{Improving Over The Born Approximation: Third-Body Rejection}\label{sec:improvingBA}

In the following we wish to focus on the density range $1.5\times 10^{17}$ $\mathrm{cm}^{-3}$ $ \stackrel{<}{\sim} n_{\mathrm{e}} \leq 2.5\times 10^{17}$  $\mathrm{cm}^{-3}$, 
where the RPA is appropriate as $r_s\stackrel{<}{\sim}1$, see Ref.~\onlinecite{marchetti2014}.
In this intermediate density range, according to the discussion at the end of the previous section, and to our previous statistical analysis 
of e-e scattering \cite{marchetti2014}, we  expect the Born approximation to overestimate the e-e scattering cross-section. 

The key issue is that the standard  theory of scattering assumes that each collision involves only two bodies, 
while an overestimate of the cross-section, e.g., due to the BA, as in this case, 
increases the probability of having a third electron within the scattering cross-section. This increased probability of
a three-body scattering event can be understood given the 
usual geometrical interpretation of the cross-section \footnote{We can think the total cross-section as an effective
area (geometrical cross-section) of a sphere seen by the colliding carrier at a given energy. Notice that the total cross-section may vary greatly even in
a small energy range as indeed  happens for a Yukawa potential.}. Furthermore the problem is getting worse when
the electronic density $n_{\mathrm{e}}$ increases, as consequently the number of  scattering centres increases as well.

The presence of this \textquoteleft third body \textquoteright \, is then not accounted for within the present formalism or, 
equivalently, the overestimation of the cross-section due to BA invalidates the use of the standard theory of scattering. We wish 
then to find a method to improve over BA but that is also  easily implementable within EMC simulation techniques.
 
\begin{figure}

\begin{center}
\includegraphics*[scale=0.6]{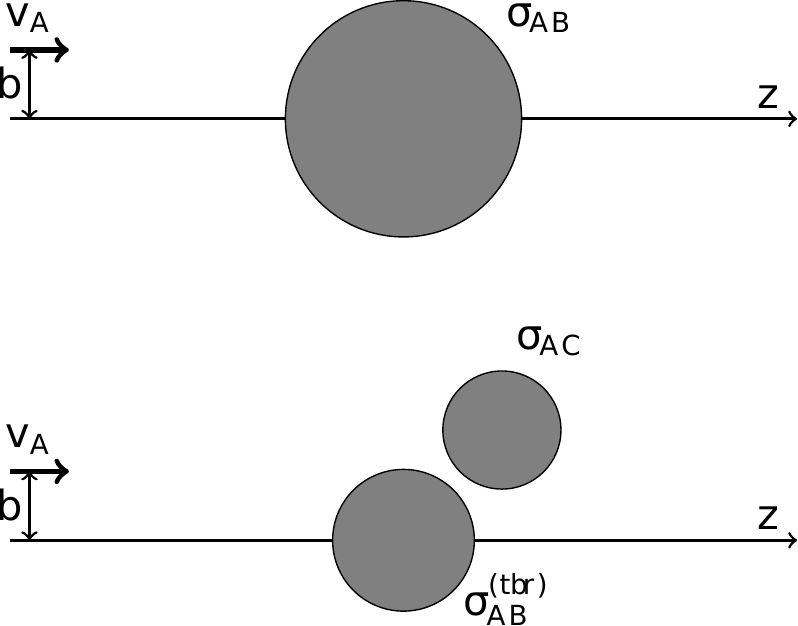}
\end{center}

\caption[] {(Top) Sketch of the cross-section for a collisional event between two particles $A$ and $B$.
(Bottom) As above but in the presence of a third body $C$. }
\label{fig:sketch}
\end{figure}

In order to retain the physical picture of a two-body collision for the scattering of an electron with an impurity, and even 
in the presence of a third carrier, Ridley devised a  method called \textquoteleft statistical screening\textquoteright \, or  
\textquoteleft third-body rejection\textquoteright \cite{ridley2000}.
 Let us use $\mathrm{b}$ to denote the impact parameter (see Fig. \ref{fig:sketch}); then the probability that there is no scattering centre 
 with impact parameter smaller than $\mathrm{b}$ is defined by \cite{ridley2000}
\begin{equation}\label{P}
P(\mathrm{b})=e^{-\pi n_{\mathrm{e}} \mathrm{a} \mathrm{b}^{2}} \, ,
\end{equation}
where $\mathrm{a}=(4 r_s \mathrm{a}^\ast_{\mathrm{B}}/\pi)\sqrt[3]{4\pi/3}$  is approximately the average distance between   the scattering centres.
Using (\ref{P}), Ridley defines the  \textquoteleft  third-body\textquoteright \, differential cross-section $ \sigma^{(\mathrm{tbr})}_{\mathrm{AB}}(\theta)$ 
($\theta$ is the scattering angle), 
 as \cite{ridley2000}

\begin{equation}
\label{eq:third_cross}
 \sigma^{(\mathrm{tbr})}_{\mathrm{AB}}\left(\theta \right)= \sigma^{Y}_{\mathrm{AB}}\left(\theta \right) \exp(-\pi n_{\mathrm{e}} \mathrm{a} \mathrm{b}^{2}) \, ,
\end{equation}
where $\sigma^{Y}_{\mathrm{AB}}(\theta)$ is the Yukawa differential cross-section.

We can give a physical interpretation of $\sigma^{\left(tbr \right)}_{\mathrm{AB}}$. In an ideal collision only two particles $A$ and $B$ are involved.
Then the total cross-section $\sigma_{\mathrm{AB}}$ is related to the probability that this collision happens. For a given energy of the 
colliding particle $A$, the effective area determined by  $\sigma_{\mathrm{AB}}$ can be roughly thought to be a measure of the tendency 
of $A$ and $B$ to interact \cite{joachain1987}, see top of Fig. \ref{fig:sketch}.

Let us now add a third-body $C$ to the system, see bottom figure in Fig. \ref{fig:sketch}. Clearly the presence
of a new potential scattering centre $C$ for the colliding particle $A$ affects its collision with $B$ and hence reduces the subsystem 
$A+B$ scattering probability. According to
Ridley's formula, Eq. (\ref{eq:third_cross}), this effect is taken into account by a reduction in the scattering probability
for the the subsystem $A + B$. Thus the total cross-section  $\sigma_{AB}$ is now given by $\sigma^{(tbr)}_{AB}$ 
through the Eq. (\ref{eq:third_cross}) integrated over the scattering
angle interval $\left[0,\pi \right]$, as  colliding particles are considered  distinguishable.
We note that in the EMC method the carriers are treated as semiclassical, distinguishable particles.  
Therefore we shall employ Ridley's method for correcting the scattering in the electron-electron collisions.

\subsection{Embedding the Third-Body Rejection Method in EMC Simulations} \label{sec:third-body method}

For the third-body rejection method, the scattering rate $\widetilde{w}_{\mathrm{ee}}^{\left(\mathrm{tbr}\right)}$ was derived by Van de Roer and  Widdershoven  
\cite{vanderoer1985} as

\begin{equation}
 \label{eq:tb_rate}
 \widetilde{w}_{\mathrm{ee}}^{\left( \mathrm{tbr} \right)}\left( v \right)= \frac{v}{\mathrm{a}}\left[1- \exp \left(-\frac{\mathrm{a} w_{\mathrm{ee}}\left(v \right)}{v}\right) \right] \, ,
\end{equation}
where $v$ (group velocity) is the speed associated to the relative motion of the colliding carriers.
We shall use  Eq. (\ref{eq:tb_rate}) for the e-e scattering rate in order  to take into account the presence of a third carrier 
\footnote{Note that Eq. (\ref{eq:tb_rate}) does no longer depend on the impact parameter $\mathrm{b}$: in the 
EMC scattering rates there is no angular dependence and consequently no dependence upon the impact parameter.}. 
Notice also that Eq. (\ref{eq:tb_rate}) implies that $\widetilde{w}_{\mathrm{ee}}^{\left( \mathrm{tbr} \right)}$ is always smaller than
$w_{\mathrm{ee}}$.

In the following we are guided by analogy and computational simplicity; a direct implementation of Eq.~(\ref{eq:tb_rate} in EMC 
would be very computationally expensive \footnote{The third-body rejection method is already implemented in several EMC simulation programs, e.g.  
IBM's DAMOCLES  and UIUC's Moka.}. Firstly as typical in the EMC method, we consider the e-e collisions as
independent from the other scattering mechanisms. Because we study the electron system at equilibrium,
we can consider the average properties of a typical carrier. Then we can interpret $\mathrm{a}/v$  as a typical  time of free flight, 
which in EMC is given by the inverse of the (average) scattering rate  $w_{\mathrm{ee}}\left(v\right)$. Then in analogy to Eq. (\ref{eq:third_cross}), we 
propose to substitute the  probability   $w_{ee}\left(v\right)$ for  $\mathrm{a}/v$, obtaining the following relation in  the carrier's relative energy
$E_{\mathrm{rel}}$
\begin{equation}
\label{eq:tb_rate1}
 w_{\mathrm{ee}}^{\left( \mathrm{tbr} \right)} \left(E_{\mathrm{rel}}\right)= w_{\mathrm{ee}}\left(E_{\mathrm{rel}}\right)\left[1- e^{ \left(-\frac{\mathrm{a} m^{\ast} w_{\mathrm{ee}}\left(E_{\mathrm{rel}}\right)}{\sqrt{2 E_{\mathrm{rel}}}}\right)} \right] \, .
\end{equation}

Consistent with Eq. (\ref{eq:third_cross}), the tbr scattering probability includes now a weighting term (the term in brackets).
Moreover this algorithm  is easily implemented in our code because it introduces a mere flag in our EMC electron-electron algorithm.

\begin{figure}

\begin{center}
\includegraphics*[scale=0.7]{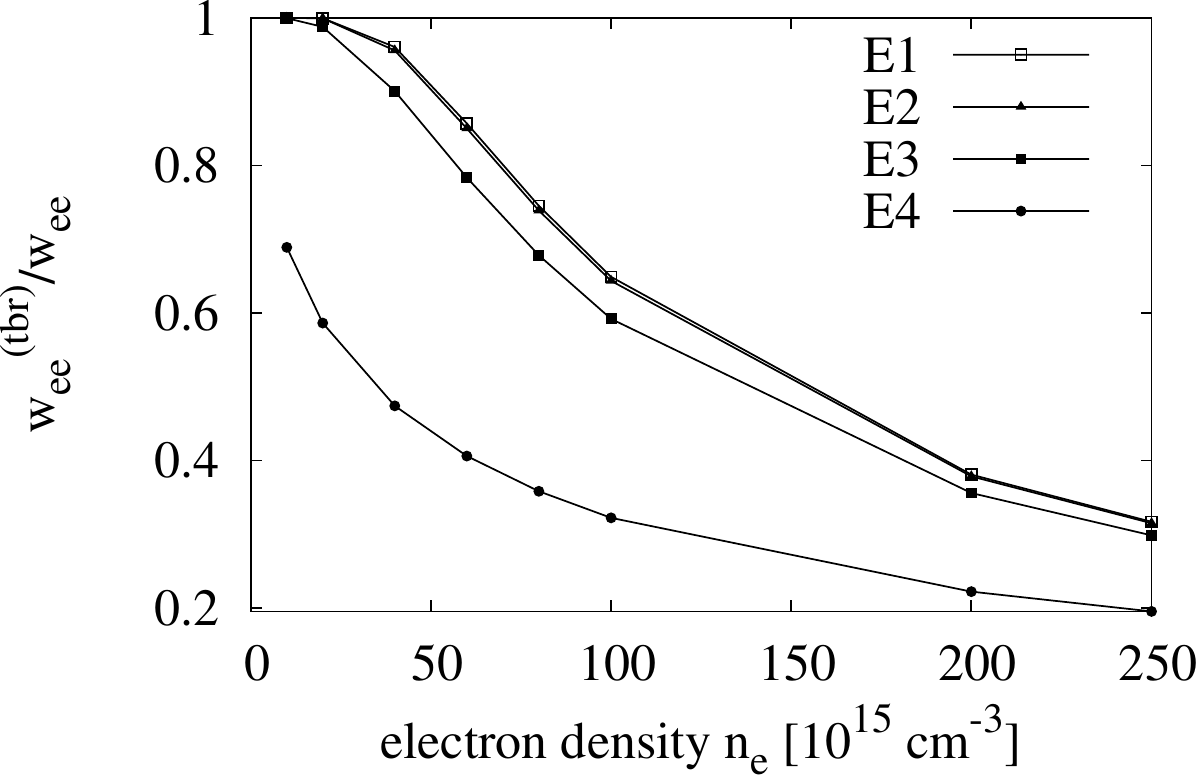}
\end{center}

\caption[] {The ratio $w_{\mathrm{ee}}^{\left( tbr \right)}/w_{\mathrm{ee}}$ calculated from Eq. (\ref{eq:tb_rate1}) against
the electron density at room temperature for four  values of the energy:  $E1=10^{-3}\times E_{\mathrm{th}}$,
$E2=10^{-2}\times E_{\mathrm{th}}$,  $E3=10^{-1}\times E_{\mathrm{th}}$ and $E4=E_{\mathrm{th}}$. $E_{\mathrm{th}}$ is the thermal energy associated to the relative motion.}
\label{fig:comparison}
\end{figure}

From Eq. (\ref{eq:tb_rate1}) we expect that $w_{\mathrm{ee}}^{\left( \mathrm{tbr} \right)}$ diminishes the e-e scattering rate with respect $w_{\mathrm{ee}}$.
In order to ascertain how much this reduction is, in Fig. \ref{fig:comparison} we plot the ratio  $w_{\mathrm{ee}}^{\left( tbr \right)}/w_{\mathrm{ee}}$ 
against the electron density $n_\mathrm{e}$ at temperature $\mathrm{T}=300$ $\mathrm{K}$ for different energies associated to the relative motion of the colliding electrons. 
Here $E_{\mathrm{th}}=(3/4)k_{\mathrm{B}}\mathrm{T}\sim 18.75$ $\mathrm{meV}$ is the electron thermal energy associated to the relative motion.

In Fig. \ref{fig:comparison} it is evident that the tbr method reduces the e-e scattering rate strongly at higher electronic concentrations.
For the thermal  carriers in the density range $n_e=1.5 \cdot 10^{17}$ to $ 2.5\cdot 10^{17}$ $\mathrm{cm}^{-3}$ the reduction is about $70\%$ to  $80\%$ 
while it is $50\%$ to  $70\%$  for energies  smaller than $0.1 E_{\mathrm{th}}$.


Notice that, for the electron densities  $n_e\stackrel{<}{\sim}1.5\times 10^{17}~ \mathrm{cm}^{-3}$, the RPA starts to break down \cite{marchetti2014}.
This implies that the electron-electron interaction  may no longer be considered as  
 a Yukawa-type potential and therefore may  fail to model the actual interelectronic potential in that doping region. In this regime the use of the tbr 
 method would not be justified.

\section{Comparison With Experiments}\label{sec:results}

In this section we present  our numerical results for the spin relaxation time $\tau_{\mathrm{s}}$ , and compare them to the available experimental data.

Apart from assuming an exponential decay of the total spin polarization in the $z$-direction, we note that our simulations have \emph{no fitting parameters}. 
In particular the spin orbit coupling value used is {\it not} fitted, but we use the value suggested
by Oertel et al. \cite{oertel2008} for their experimental data: $\gamma_{\mathrm{so}}=21.9$ $\mathrm{eV}$ {\AA}$^{3}$ .

\begin{figure}
\begin{center}
\includegraphics*[scale=0.7]{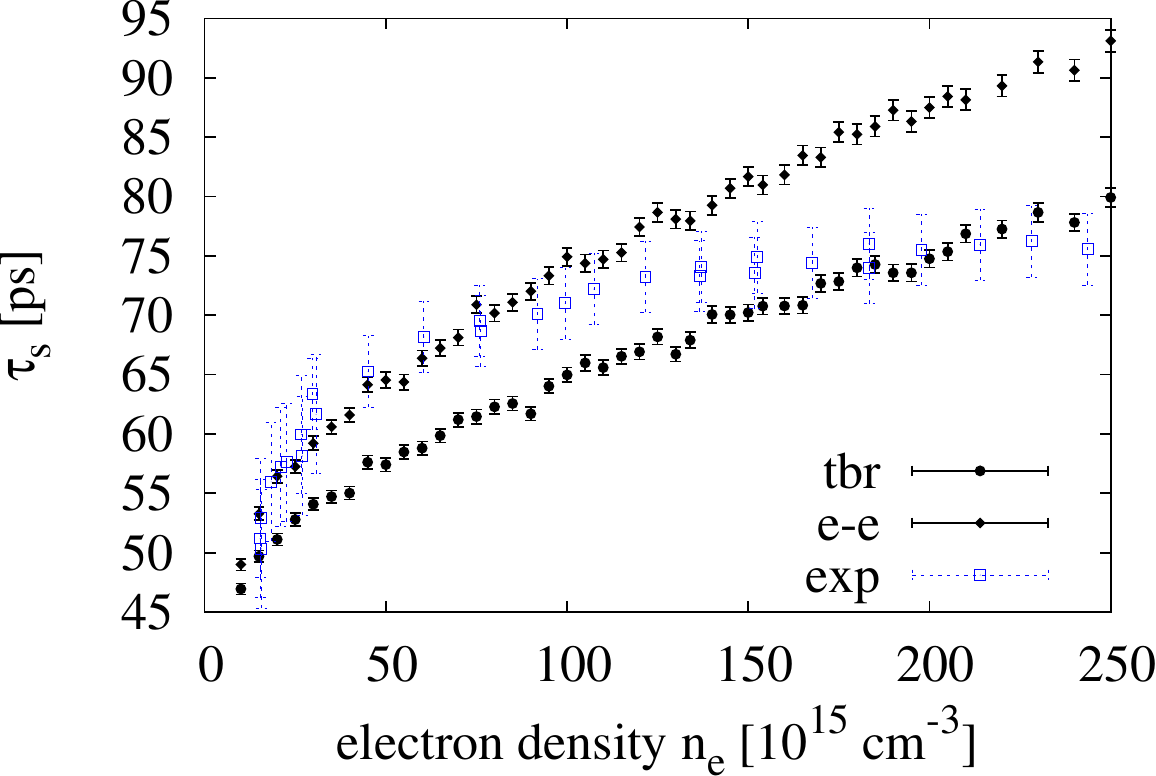}
\end{center}

\caption[] {Spin relaxation time versus electron density. All data include e-e scattering, with (solid circles), 
and without (solid diamonds) third-body rejection corrections. Parameters:
$N= 25,000$ , $T = 300$ $\mathrm{K}$, and $\gamma_{\mathrm{so}}=21.9$ $\mathrm{eV}$ \AA $^{3}$. The experimental data
from Ref.~\onlinecite{oertel2008} are plotted as well (empty square symbols).}
\label{fig:tvsdensity}
\end{figure}

In Ref.~\onlinecite{marchetti2014} we observed that, when we include e-e interaction as described in Sec. \ref{sec:model}, our results for densities
$10^{16} \mathrm{cm}^{-3} \stackrel{<}{\sim}  n_{\mathrm{e}} \stackrel{<}{\sim} 10^{17} \mathrm{cm} ^{-3}$ are in {\it very good agreement with the experimental data} (empty square symbols)
reproduced in Fig. \ref{fig:tvsdensity}. In Ref.~ \onlinecite{marchetti2014} we interpreted these results as an accidental cancellation of the effects
of the Born approximation, see Sec.~\ref{sec:model},  and the use of the Yukawa potential, i.e. overestimation and underestimation of the e-e scattering
contribution to the motional narrowing effect \cite{fabian2007} respectively.

 However, at higher densities, our results from Ref.~\onlinecite{marchetti2014} (diamonds, Fig. \ref{fig:tvsdensity}) start to 
 overestimate the experimental data for $\tau_{\mathrm{s}}$, reaching $\sim 20\%$ 
 overestimate when $n_{\mathrm{e}} = 2.5\times 10^{17}  \mathrm{cm}^{-3}$.

Let us assume for a moment  that all the other scattering mechanisms give a reasonable contribution to the motional narrowing (we shall return to this
issue  in Sec.\ref{sec:Discussion}). We  speculate that the overestimate of $\tau_\mathrm{s}$ for $n_{\mathrm{e}}\stackrel{>}{\sim} 10^{17}  \mathrm{cm}^{-3}$ is again
an effect of the BA  we employ for the e-e scattering, but now  the validity of RPA ( $r_{\mathrm{s}} \lesssim  1$) \cite{giuliani2005} in this electron
density region says that in general the Yukawa potential is more suitable for modeling the e-e interaction, so no accidental cancellation
is possible. In the following we shall then check 
whether including corrections due to third-body-rejection to the e-e interaction  improves our results.

In  Fig. \ref{fig:tvsdensity}  we compare the experimental results from Ref.~\onlinecite{oertel2008} with our calculations of the spin relaxation times 
 including e-e interactions with ($\tau_{\mathrm{s}}^{\mathrm{ee},\mathrm{tbr}}$) 
and without ($\tau_{\mathrm{s}}^{\mathrm{ee}}$) corrections due to third-body rejection. Following the trend in Fig. \ref{fig:comparison},
the reduction of the e-e scattering rate due to the inclusion 
of  third-body rejection causes a  reduction of the spin relaxation times at all densities, which, for the range of densities studied, 
becomes more significant with increasing density.

Looking at Fig. \ref{fig:tvsdensity}, we observe that, at relatively high densities, when the use of the tbr technique is justified, $\tau_{\mathrm{s}}^{ee,tbr}$ significantly 
improves over $\tau_{\mathrm{s}}^{\mathrm{ee}}$,  giving {\it results within the experimental error bars} for $1.5\times 10^{17}  \stackrel{<}{\sim}  n_{\mathrm{e}}  \stackrel{<}{\sim}  2.5 \times 10^{17}$. 

However for decreasing densities we see that
$\tau_{\mathrm{s}}^{\mathrm{ee},\mathrm{tbr}}$ departs from the experimental results, underestimating $\tau_{\mathrm{s}}$, and significantly so for the density range where 
$\tau_{\mathrm{s}}^{\mathrm{ee}}$ has a good agreement with the experimental data and the use of tbr is \emph{not} justified.

\section{Discussion}\label{sec:Discussion}

In this section we discuss our numerical results in the region  $n_\mathrm{e} \stackrel{>}{\sim} 1.5 \times 10^{17}  \mathrm{cm}^{-3}$ where we expect
that our model  is reasonably closer to the real physical system.

All the scattering mechanisms we included in our calculations, contribute to the motional narrowing effect, and in turn to the spin relaxation times 
$\tau_{\mathrm{s}}$. The 
electron-phonon interactions which we consider unscreened \cite{pugnet1981},  are given by the electron-LO scattering and the electron-LA scattering. 
Both are correctly estimated; in fact the scattering rates we use are well established in the literature and have given quantitatively accurate results
in studies of charge transport \cite{jacoboni1989,ridley2000}. Furthermore the electron-LA scattering  is negligible at room temperature.

The e-i scattering, similarly to the e-e scattering, could be expected to be, due to the low energies involved, also beset by the Born approximation. 
Moreover given that $n_{\mathrm{i}}=n_{\mathrm{e}}$, $n_{\mathrm{i}}$ being the doping concentration, one might expect that there is also the need for
the tbr correction of the e-i scattering; indeed this method was firstly applied to the electron-impurity collisions \cite{ridley2000}.

In the following we shall explain why this is  not the case. Discarding all the other scattering mechanisms
Meyer and Bartoli performed the phase-shift calculations \footnote{ Mayer-Bartoli's
phase-shift analysis has also been implemented in IBM's DAMOCLES program. }of the e-i scattering contribution to the electron mobility for {\it n}-type GaAs and then compared it to 
the mobility obtained using the e-i scattering in BA (Brooks-Herring approach) \cite{meyer1981}. At room temperature
in the doped region of interested, they found a very good agreement between
the two methods. This implies that we can safely assume that the e-i scattering is quite accurate 
when estimated in BA approximation and for the densities of interest. From this result, recalling that the tbr method would diminish 
the scattering probability (see Eq.  (\ref{eq:third_cross})
or equivalently Eq. (\ref{eq:tb_rate})), it is evident that the tbr correction is unnecessary for e-i scattering. This is confirmed a posteriori by
the good agreement of our calculations with the experimental data, as shown in Fig. \ref{fig:tvsdensity}. 

\begin{figure}
\begin{center}
\includegraphics*[scale=0.7]{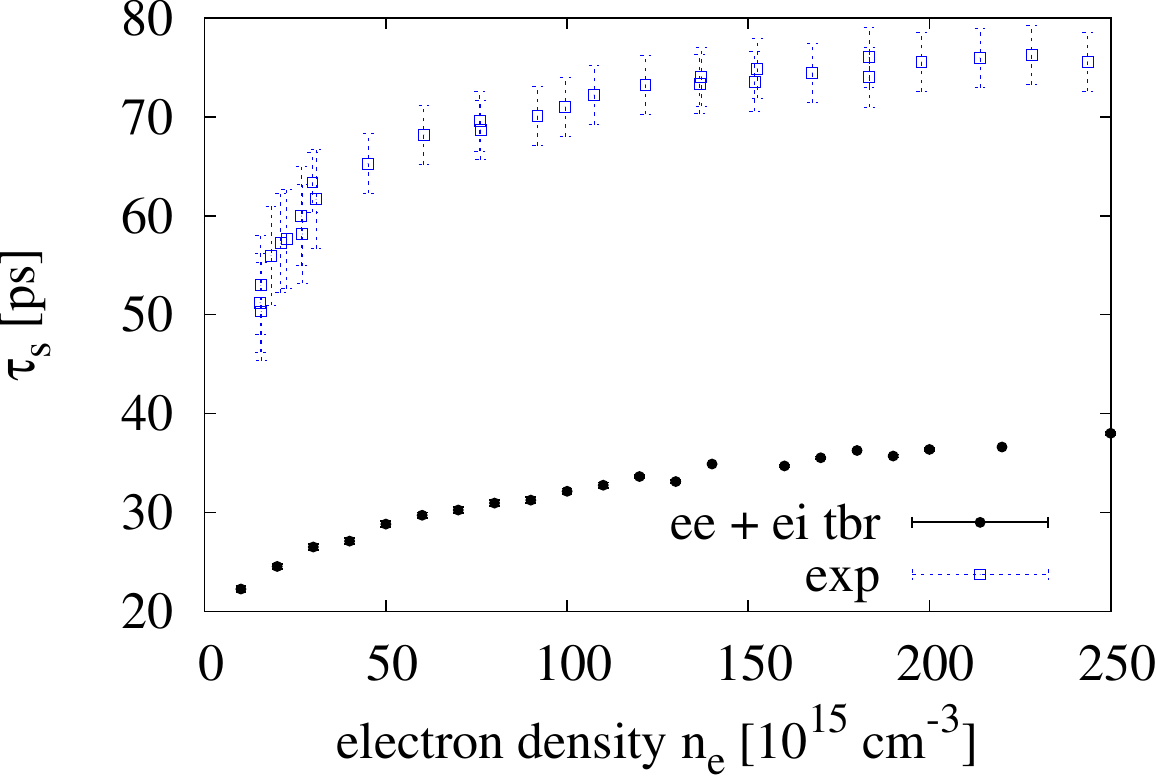}
\end{center}

\caption[] {Spin relaxation time versus electron density. The curve with solid circles represent
the data from simulations when the tbr method is applied to both electron-electron and
electron-impurity scattering. Parameters:
$N= 25,000$, $T = 300$ K, and $\gamma_{so}=21.9$ eV \AA $^{3}$. The experimental data
from Ref.~\onlinecite{oertel2008} are plotted as well (empty square symbols).}
\label{fig:tvsdensity_tbr}
\end{figure}

We note that if we apply the tbr method to
both e-e and e-i collisions, we are no longer able to reproduce the experimental data from our simulations, see Fig. \ref{fig:tvsdensity_tbr}.
Now  the spin relaxation times (solid circles)  are  understimated to a great degree
(more than $50\%$ in some cases).
These data show that the overall amount of the Coulomb scattering rate in
our numerical simulations, is  also decreased as we recover results similar to the case in which e-e interactions are completely
neglected, see Ref.~\onlinecite{marchetti2014}. Notice that the  collisional energy involved in e-e collisions is smaller 
 than the one available in the electron-impurity scattering processes because  part of the energy 
is absorbed by the center of mass's motion
\footnote{If $E_{\mathrm{lab}}$ denotes 
the energy of an electron $A$ in the laboratory frame of reference colliding with an electron $B$, then the energy $E_{rel}$, available for collision 
in the center of mass  frame of reference, is given by $E_{rel}=m_B/\left[\left(m_A+m_B\right)E_{lab}\right]$,
i.e., $E_{rel}=0.5 E_{lab}$ for the case of the
e-e scattering \cite{schiff1968}. Notice that the two colliding carrier  masses may be slighlty different due to the  nonparabolicity of the conduction band. However
this has in general a negligible effect on the previous consideration.}.
This small collisional energy difference between the Coulomb scattering processes may be then  the main cause of the overstimate for the spin relaxation
times obtained from our calculations due to the electron-electron scattering in Born approximation.

The close relation between the motional 
narrowing phenomenon and the collisional regime, in addition with the good agreement of our numerical results
via the third-body rejection method applied to the electron-electron scattering only, contributes to 
confirm that the e-e collision rate is overstimated in BA. 
A quantitative analysis of the overestimate can be achieved only evaluating the phase-shifts of  the electron-electron
scattering for the parameter set of interest, i.e., $n_{\mathrm{e}}$, $\mathrm{T}$, $ \beta_{\mathrm{TF}}$ and
the collisional energy range of the carriers involved in the simulations.

We wish  to make some additional remarks on the  limits of the physical model we adopt 
in  the present work. First of all, the finite temperature LTFA we use to estimate the screening effects is consistent with 
a static interaction. This is the case for the electron-impurity interaction where the center of mass of
the two-body system can be considered at rest. In an electron-electron interaction because  
the center of mass moves with speed $v_{c.m}$ in the dielectric medium, the screening has also a frequency dependence
$\omega=\mathbf{q}\cdot \mathbf{v_{c.m}}$ where $\mathbf{q}$ is the momentum transfer \cite{meyer1983}. 
Then the current results could be improved by including the frequency dependent dielectric function 
 $\epsilon(q,\omega=\mathbf{q}\cdot \mathbf{v_{c.m}},\mathrm{T})$ \, \cite{giuliani2005}.

Also we conjecture that  quantum intereference in the electron-electron collisions starts to play a role in
this regime. Indeed  two colliding electrons are in principle indistinguisable fermions. The signature of 
these effects can be inferred from the Fermi temperature $\mathrm{T}_\mathrm{F}$ which for the region of interest roughly varies from $178$ $\mathrm{K}$ to $250$ $\mathrm{K}$; a 
comparison of these values with $\mathrm{T}=300$ $\mathrm{K}$ shows that our system is in an intermediate regime, $\mathrm{T}\stackrel{>}{\sim}\mathrm{T}_{\mathrm{F}}$, and therefore 
neglecting the antisymmetry of the colliding electrons's wavefunction, which must include triplet and singlet spin states, might affects our numerical results.

This fact might then explain the flattening of the experimental data in Fig. \ref{fig:tvsdensity}, while the monotonically increasing  behaviour
of the curves $\tau_\mathrm{s}^{\mathrm{ee}}$ and  $\tau_{\mathrm{s}}^{\mathrm{ee},\mathrm{tbr}}$ obtained from our calculations is instead consistent with a nondegenerate regime
($ \mathrm{T}\gg \mathrm{T}_\mathrm{F}$)  where the electron-electron
scattering rate $w_{\mathrm{ee}} \propto n_\mathrm{e}$ \cite{jiang2009}$^{,}$ \footnote{Notice  that the e-e scattering formula we use \cite{marchetti2014}, has a slightly
more complicated dependence upon the electron density, as $n_e$ enters also into the variable $\beta_{\mathrm{TF}}$. However in first
approximation we can assume that $w_{\mathrm{ee}} \propto n_{\mathrm{e}}$.}.
Therefore we should expect that the quantum mechanical interference due to the fermionic nature of the carriers, diminishes the
e-e scattering contribution to the spin relaxation times $\tau_\mathrm{s}$ when $n_\mathrm{e}$ is relatively high. Anyway the system
is far from the degenerate regime in which Pauli principle
dominates making the electron-electron collision negligible, as it is typical in the metals.
Instead what we guess is happening in the  physical system, is analogue to what one usually observes, for instance, in the angular distribution of
$\alpha - \alpha$ scattering at relatively low energy $\sim 150$ $\mathrm{KeV}$, where the quantum interference diminishes the 
Coulomb scattering for certain values of the scattering angles \cite{heydenburg1956}.

\section{Conclusions}

We have shown that the third-body rejection method can be  successfully employed for studying  spin decoherence in semiconductors. With this
tool we have been able to obtain a good agreement of our calculations with the experimental data in the doping region where 
its use can be justified on a physical basis. The third-body method handled quite well the  intertwined effects
of the Born approximation and  a multiple scattering regime. 

This little known technique was already successfully employed in the calculations of the mobility in the case of the electron-impurity scattering.
Thus we think that this method, based upon a simple physical insight, deserves more attention, and in particular a better understanding
of its relation with the Born approximation. 

Further work  could include the study of quantum intereference  due to the direct and exchange transitions \cite{moskova1994}
of the colliding electrons.
This could be done, for instance, using the non-relativistic Mott scattering formula \footnote{With regards to this formula \cite{joachain1987}, 
we  recall that 
it describes the unpolarized scattering probability of fermions, averaging over the possible triplet and singlet spin states
which they can form. Then some care should be taken to implement this condition.}.

Additionally the dynamical screening \cite{meyer1983} and
nonparabolicity of the conduction band \cite{jacoboni1989} could be included in the model; however we suspect that the 
relative improvements might be largely shadowed by the failing of the Born approximation, the constraint on 
which the ensemble Monte Carlo method strongly relies.

In this respect the EMC method, through the  calculations of the spin relaxation
times, can give a test bed for modeling the many-body interactions in a semiconductor.
Our simulations have demonstrated a very high sensitivity of the spin dephasing to the accuracy of 
the modeling of the electron-electron interactions.

\begin{acknowledgments}
We acknowledge support from EPSRC Grant
No. EP/F016719/1. We wish to thank the reviewer for pointing out some important references of the vast EMC literature.
\end{acknowledgments}

\bibliography{jap_references}{}
\bibliographystyle{ieeetr}

\end{document}